\documentclass[%
 reprint,
superscriptaddress,
groupedaddress,
preprintnumbers,
nofootinbib,
 amsmath,amssymb,
 aps,
dvipsnames
]{revtex4-2}
\pdfoutput=1

\usepackage{graphicx}
\usepackage{dcolumn}
\usepackage{bm}
\usepackage[dvipsnames]{xcolor}
\usepackage[normalem]{ulem} 
\usepackage{tikz}
\usepackage{bbm}
\usetikzlibrary{decorations.pathmorphing}
\usetikzlibrary{decorations.markings}

\newcommand{\Nm}{ {\textrm{Fr}}}
\newcommand{\Wf}{ W_{\Nm}}
\newcommand{\Pf}{ \Pi_{\Nm}}
\newcommand{\Dim}{ \textrm{d} }
\newcommand{\Dyn}{ \textrm{I} }

\begin{document}

\preprint{IPPP/24/13}

\title{Fractional-charge hadrons and leptons to tell the Standard Model group apart}

\author{R. Alonso}

\author{D. Dimakou}

\author{M. West}
\affiliation{Institute for Particle Physics Phenomenology, Durham University, \\Durham, DH1 3LE, United Kingdom }
\begin{abstract}
The gauge group of strong and electroweak interactions in Nature could be any of the four that share the same Lie algebra,  $SU(3)_c\times SU(2)_L\times U(1)_Y/Z_p\equiv G_p$
with $Z_p=\left\{Z_6,Z_3,Z_2,Z_1\right\}$. 
Each of these cases allows in its spectrum for the matter fields of the SM but also for new distinctive representations, e.g. under the assumption that $q_L$ possesses the minimum possible hypercharge in Nature, $G_p$ allows for particles with a multiple of $p\,e/6$  for electric charge. This letter discusses how these new possibilities in the spectrum could be used to tell the SM group apart.
\end{abstract}

\maketitle

\section{Introduction}
Group theory and symmetry permeate all of particle physics and provide the foundation for our most fundamental theory of nature. The group \begin{align}\tilde G=SU(3)_c\times SU(2)_L\times U(1)_Y\,,\label{eqtG}\end{align} and its six elementary representations to be found in Nature (and in tab.~\ref{tab:chargesSM}) are often showcased as the minimum set of mathematical constructs that suffice to explain the majority of our experience of the universe, gravity excluded. The simplicity of the theory that arose from the distilled experimental effort of many decades is a most remarkable fact indeed, yet such an exhibit should come with a note: $\tilde G$ is one of the four groups compatible with observation~\cite{Hucks:1990nw,Tong:2017oea}. Three other compact groups share the same Lie algebra of $\tilde G$ and lead to the same perturbative dynamics yet are different in their global structure. The reader might be tempted, having read this far, to dismiss the note and discussion of groups as purely academic; it is not so. As pointed out in~\cite{Tong:2017oea}  and elaborated in this letter, the discovery of new fractionally charged particles will help us tell apart the true group of gauge interactions.
\begin{table}[h!]
    \centering
\begin{tabular}{c|c|c|c|c|c|c}
     & $q_L$& $u_R$& $d_R$ & $\ell_L$ & $e_R$ & $H$\\ \hline
    $ U(1)_Y$ & $1/6$ & $2/3$ &$-1/3$ &$-1/2$ &$-1$&$1/2$ \\
    $SU(2)_L$ &$2$ & $1$ & $1$& $2$ & $1$&2\\
    $SU(3)_c$ & $3$ & $3$ & $3$ & $1$ &$1$& $1$
\end{tabular}
\caption{Table with SM representations and our convention for hypercharge. When conveying the same information, not in table format, we will have that e.g. $q_L$ is  in $(3,2)_{1/6}$.}
    \label{tab:chargesSM}
\end{table}

Fractional-charge particles themselves have been discussed in the literature and searched for experimentally for many decades.
They appear in embeddings of GUTs~\cite{Gell-Mann:1976xin} and in string theory realisations~\cite{Wen:1985qj,Athanasiu:1988uj}, and their phenomenology was studied and put against experiment~\cite{Perl:2009zz}, while cosmology sets stringent constraints if the universe ever got hot enough for them to thermalise~\cite{Langacker:2011db}. The current theory approach offers a different light on these theories~\cite{Aharony:2013hda,Tong:2017oea,Davighi:2019rcd,Wan:2019gqr}, but it also breaks from the background in which fractional-charge particles emerged to reduce the problem to its essential components with no reference to theories beyond the SM.

The structure of this letter is as follows: sec.~\ref{SMGr} introduces the concept of locally identical but globally different groups to apply it to the SM group, sec.~\ref{secSpct} discusses the different representations for each group and derives the quantisation conditions on electric charge while sec.~\ref{Sec:Smaller} studies the phenomenology of fractionally charged particles. A summary is to be found in sec.~\ref{secSum} while additional results are placed in~\ref{secApp}.

\section{The Gauge Groups of the Standard Model} \label{SMGr}

The Lie algebra that characterises the interactions of charged particles and mediators in a gauge theory and determines the perturbative $S$-matrix is, in general, shared by several groups. These groups differ by whether the zentrum or centre $Z$, i.e. the set of elements that commute with every other element, is present or partially absent.

{$\mathbf{SU(2)}$ {\bf vs} $\mathbf{SO(3)}$.} An illustrative example is the case of $SU(2)$ and $SO(3)$. Both groups are locally the same and so the generators of each satisfy
\begin{align}
 SU(2):&  &  \left[\frac{\sigma_i}{2}\,,\frac{\sigma_j}{2}\right]&=i\epsilon_{ijk}\frac{\sigma_k}{2}\,,\label{SU2F}\\
 SO(3):&  &  \left[T_i\,,T_j\right]&=i\epsilon_{ijk}T_k\,,\label{SO3F}
\end{align}
with $\sigma_i$ the Pauli matrices and $T_i$ the hermitian anti-symmetric matrices in 3 dimensions $T_i^T=-T_i$ with tr$(T_iT_j)=2\delta_{ij}$. The crucial difference lies in the centre, $SU(2)$ has $Z(SU(2))=Z_2=\left\{1,\xi\right\}$ with the element $\xi$ in the fundamental representation being $-\mathbbm{1}$. On the other hand $SO(3)$ has no non-trivial centre, $Z(SO(3))=1$ (i.e. the element $\xi$ is missing); if one sets off from the identity in the same direction in both groups, say Exp$(i\alpha \sigma_3/2)$ in $SU(2)$ and Exp$(i\alpha T_3)$ in $SO(3)$, one would reach $\xi$ for $\alpha=2\pi$ in $SU(2)$ or circle back to the identity for $SO(3)$. One might then be tempted to write:
\begin{align}
    SO(3)=\frac{SU(2)}{Z_2}\,,
\end{align}
i.e. the quotient of $SU(2)$ by $Z_2$, and one would be right. The application of one such expression is as follows: take the numerator, $SU(2)$, and remove the centre by keeping only those representations $R$ that do not `see' it, i.e. $\xi R=R$ and so in this case the fundamental of $SU(2)$ in eq.~(\ref{SU2F}) is discarded and the Pauli matrices as generators with it. For this letter, the lesson to be taken away is that {\it taking the quotient restricts the possible representations and one does obtain a different group.}

{$\mathbf{U(1)\times SU(N)}$ {\bf vs} $\mathbf{U(N)}$.} One more example to gear up to the Standard Model case is $U(N)$ and $U(1)\times SU(N)$. Here we can use intuition first and then connect to the centre discussion.  
Take the action on a fundamental representation $F$
\begin{align}
    F\to e^{i\theta_0 Q} e^{iT_a\theta^a}F=e^{i\theta_0 Q_F} e^{iT_a\theta^a}F\,,
\end{align}
where $\theta_0,\theta_a$ are $N^2$ real group parameters, $T_a$ the $SU(N)$ generators, $Q$ the $U(1)$ charge operator and $Q_F$ the charge of $F$. There is some arbitrariness in the charge $Q_F$ definition; what are not arbitrary however are the charge ratios. For $U(N)$ one has that $U(1)$ and $SU(N)$ actions are tied in therefore one can obtain any other representation combining fundamental representations. An example is a $SU(N)$ singlet with $U(1)$ charge obtained by taking the asymmetric combination of $N$ fundamentals which would transform as
\begin{align}
    \det(F\,,\dots \,,F^{(N)})\to e^{i\theta_0 NQ_F}\det(F\,,\dots\,, F^{(N)}) \,,
    \end{align}
    that is, for $U(N)$ the minimum nonzero charge of a $SU(N)$ singlet is $N$ times that of a fundamental $Q_S=NQ_F$. In contrast for $U(1)\times SU(N)$ no such correlation between $SU(N)$ and $U(1)$ charges exists. 
    
    To connect with the centre discussion we note first that for $SU(N)$ the generalisation of the $SU(2)$ result is $Z(SU(N))=Z_N$ with the first nontrivial element being $e^{2\pi n_Ni/N}$ where $n_N$ is the n-ality of the representation, a positive integer mod $N$ (1 for the fundamental, 2 for the symmetric, 0 for the adjoint etc) whereas $U(1)$ is its own centre. In the case of $U(N)$ however all these elements are not distinct, the action of $Z_N$ can be always `undone' by a $U(1)$ action with $\theta_0=-2\pi/(Q_F N)$.
    \begin{figure}
\centering
\begin{tikzpicture}
    \draw [Gray,thick,dashed,->] (0,-0.2) -- (0,4) node [anchor=east] {$\frac{Q}{Q_F}$};
    \draw [Gray,thick,dashed,->] (1,-0.2) -- (1,4);
    \draw [Gray,thick,dashed,->] (2,-0.2) -- (2,4);
    \draw [Gray,thick,dashed,->] (6.5,-0.2) -- (6.5,4);
       \draw [Gray,thick,dashed,->] (-0.2,0) -- (7,0)node [anchor=north] {$n_N$} ;     
          \filldraw [black] (0,0) node [anchor=east] {$0\,\,\,\,$} circle (2pt);
          \filldraw [black] (0,3) circle (2pt) node [anchor=east]{$N$};          
          \filldraw [black] (1,1/2) circle (2pt) node [anchor=east] {1};
                \filldraw [black] (1,3+1/2) circle (2pt) node [anchor=east] {$N\!+\!1$};
          \filldraw [black] (2,1) circle (2pt) node [anchor=east] {$2$};
          \filldraw [black] (6.5,2.5) circle (2pt) node [anchor=east] {$N-1$};
   \end{tikzpicture}
   \caption{Possible representations in the plane of n-ality vs $U(1)$ charge for $U(N)$.}\label{fig:UN}
\end{figure}
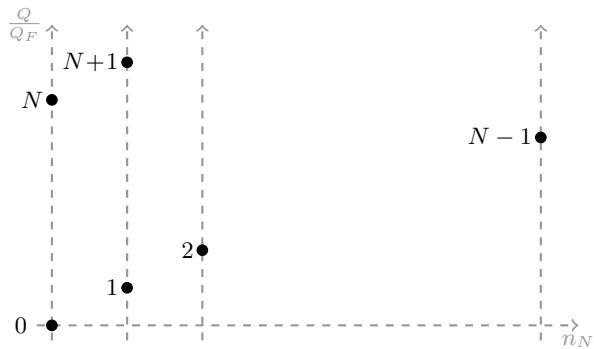
This can be put in operator form in that the element
    \begin{align}
        \xi=e^{2\pi i n_N/N} e^{-2\pi iQ/(Q_FN)}\,,\label{xiN}
    \end{align}
    acting on $U(N)$ representations returns the identity, i.e.
   \begin{align}
    U(N):& & \xi R&=R\,,\\
    &&2\pi\left(\frac{n_N(R)}{N}-\frac{Q_R}{Q_FN}\right)&=2\pi \mathbb{Z}\,.
\end{align} 
    One of the solutions to this equation returns the result derived above for $Q_S/Q_F$ in $U(N)$ yet it is clear that using the centre is the systematic way to obtain all representations, the first few shown  in fig.~\ref{fig:UN}. Lastly, the element $\xi$ generates the centre $Z_N=\left\{ 1,\xi,\xi^2,\dots,\xi^{N-1}\right\}$ and the condition of invariance under $\xi$ equates invariance under $Z_N$ (if $\xi R=R$ it follows $\xi^pR=R$ with $p$ integer) so that the equation above is the result of the relation
    \begin{align}
U(N)=\frac{U(1) \times SU(N)}{Z_N}\,.\label{UNZN}
\end{align}    
    This second example shows how taking the quotient with an abelian factor in the numerator quantises and correlates charge with $n$-ality. 
    
    Quantisation of charge however also follows in $U(1)$. Consider increasing the phase in a $U(1)$ transformation on the elementary particle $F$ until we circle back to the identity $e^{iQ_F\theta}=1$, i.e. $Q_F\theta=2\pi$. For consistency the same transformation in any other charged state with charge $Q_i$ should also be the identity so $Q_i\theta/Q_F\theta=\mathbb{Z}$. In this regard it is useful to express  $U(1)$ (a.k.a. $S^1$) as the $N\to 1$ limit of eq.~(\ref{UNZN})
\begin{align}
    U(1)=&\frac{U(1)}{Z_{1}}=\frac{\mathbb{R}}{(2\pi/Q_F)\mathbb{Z}}\,,\label{U1Q}\\
    \theta\in \mathbb{R}\,,& \quad\theta\sim \theta+\frac{2\pi\mathbb{Z}}{Q_F} \,,
\end{align}
that is: the real line with an equivalence relation. The seemingly redundant numerator on the left hand side is $Z_1=\xi(N\to 1)=e^{-2\pi Q_i/ Q_F}$ which is only truly redundant, and the equation above sensical, when equaling the identity, i.e. if charge is quantised in units of $Q_F$. In the context of $U(1)\times SU(N)$ one can understand this requirement since the zentrum $Z_N$ is present but for it to be a group it should be closed, i.e. the generating element in eq.~(\ref{xiN}) should satisfy $\xi^N=e^{-2\pi Q i/Q_F}=1$. 

\begin{figure}
    \centering
    \begin{tikzpicture}
        \draw [Gray,thick, dashed,->] (-0.2,0) -- (2.2,0) node [anchor=west] {$n$};
        \draw [Gray,thick,dashed,->] (0,-0.2) -- (0,6) node [anchor=east] {$Q$};
        \draw [Gray,thick,dashed] (-0.2,1)  node [anchor=east] {\color{black}$Q_F$} -- (2.2,1);
        \filldraw [black] (0,0) circle (2pt);
        \filldraw [blue]  (0,1) circle (2pt);
        \filldraw [blue]  (0,2) circle (2pt);
        \filldraw [black] (0,3) circle (2pt);
        \filldraw [blue] (0,4) circle (2pt);
        \filldraw [blue] (0,5) circle (2pt);
        \draw [Gray,thick,dashed,->] (1,-0.2) -- (1,6) ;
        \filldraw [ForestGreen] (1,1) circle (3pt);
        \filldraw [black] (1,4) circle (2pt);
        \filldraw [blue]  (1,0) circle (2pt);
        \filldraw [blue]  (1,2) circle (2pt);
        \filldraw [blue]  (1,3) circle (2pt);
        \filldraw [blue] (1,5) circle (2pt);
        \draw [Gray,thick,dashed,->] node [anchor=east] {\color{black}0\,\,\,\,} (2,-0.2) -- (2,6) ;
        \filldraw [black] (2,2) circle (2pt);
        \filldraw [black] (2,5) circle (2pt);
        \filldraw [blue] (2,0) circle (2pt);
        \filldraw [blue] (2,1) circle (2pt);
        \filldraw [blue] (2,3) circle (2pt);
        \filldraw [blue] (2,4) circle (2pt);
    \end{tikzpicture}\quad
    \begin{tikzpicture}
        \draw [Gray,thick, dashed,->] (-0.2,0) -- (2.2,0) node [anchor=west] {$n$};
        \draw [Gray,thick,dashed,->] (0,-0.2) -- (0,6) node [anchor=east] {$Q$};
        \draw [Gray,thick,dashed] (-0.2,4)  node [anchor=east] {\color{black}$Q_F$\,} -- (2.2,4);
        \filldraw [black] (0,0) circle (2pt);
        \filldraw [blue]  (0,1) circle (2pt);
        \filldraw [blue]  (0,2) circle (2pt);
        \filldraw [black] (0,3) circle (2pt);
        \filldraw [blue] (0,4) circle (2pt);
        \filldraw [blue] (0,5) circle (2pt);
        \draw [Gray,thick,dashed,->] (1,-0.2) -- (1,6) ;
        \filldraw [black] (1,1) circle (2pt);
        \filldraw [ForestGreen] (1,4) circle (3pt);
        \filldraw [blue]  (1,0) circle (2pt);
        \filldraw [blue]  (1,2) circle (2pt);
        \filldraw [blue]  (1,3) circle (2pt);
        \filldraw [blue] (1,5) circle (2pt);
        \draw [Gray,thick,dashed,->] node [anchor=east] {\color{black}0\,\,\,\,} (2,-0.2) -- (2,6) ;
        \filldraw [black] (2,2) circle (2pt);
        \filldraw [black] (2,5) circle (2pt);
        \filldraw [blue] (2,0) circle (2pt);
        \filldraw [blue] (2,1) circle (2pt);
        \filldraw [blue] (2,3) circle (2pt);
        \filldraw [blue] (2,4) circle (2pt);   
    \end{tikzpicture}\caption{Two lattices of representations in the charge vs triality plane for both $U(3)$ (black entries) and $U(1)\times SU(3)$ (blue and black) compatible with an experimentally observed representation marked in green.}
    \label{fig:U3}
\end{figure}
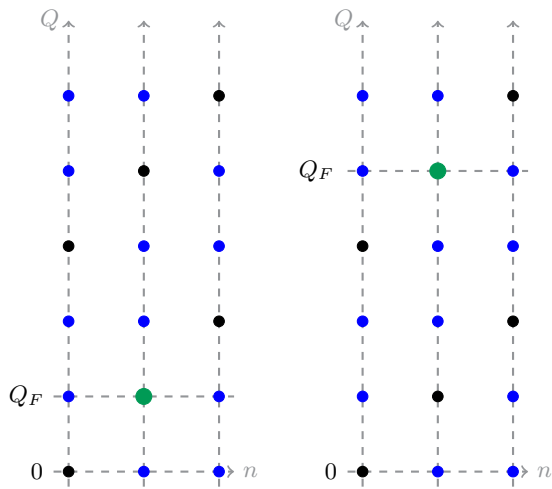

{$\mathbf{N=3}$ {\bf in the shoes of an experimentalist.}} Having outlined the differences in the spectrum of these theories, it is pertinent to turn to a thought experiment on how to determine the true gauge group. Consider an experimentalist who has discovered a fundamental of $SU(3)$ with charge $Q_F$ under an abelian group. They might have found this particle after breaking apart some bound state and observed that it behaves, insofar as they can test it, as an elementary particle, and that every other state they know of has charge an integer multiple of $Q_F$.  It is natural to then take $Q_F$ as the minimum possible quanta of charge. 
The spectrum of $U(3)$ and $U(1)\times SU(3)$ with this assumption is shown in fig~\ref{fig:U3}. The pattern for $U(3)$ charges shows correlation with the non-abelian group and if we found say a singlet with charge $Q_F$, we would have to conclude our group is $U(1)\times SU(3)$. On the other hand if one discovers a new elementary particle in a singlet representation with charge $4Q_F$, while it would be evidence for $U(3)$, one could not rule out $SU(3)\times U(1)$.

The assumption of having found the minimum quanta can be dropped or better still disproven if the experimentalist happens to discover later on another particle in the fundamental with charge a fraction of $Q_F$, let us call it $F'$ with charge $Q_{F'}$. The discussion of $U(1)$ quantisation lends itself to accommodate this possibility readily; after all who is to know one has found the smallest charge? The discussion in eq.~(\ref{U1Q}) can be changed by taking $Q_F\to Q_{F'}$ acknowledging we had misidentified the minimum quanta and instead $Q_F/Q_F'=\mathbb{Z}$. It is perhaps not so evident but equally the case that fractional charges can be accommodated in $U(3)$. Indeed one might have that the original discovery was the next-to-minimal charge for a fundamental, the entry in green in the RHS lattice of fig.~\ref{fig:U3}, which allows for a more elemental charge of $Q_F/4$. One need not stop there, the original fundamental $F$ could be two periods above the elementary charge which would therefore be $Q_F/7$ with another representation in between of charge $4Q_F/7$. We label these possibilities with a {\it compositeness degree } $k$, which is an integer that labels how many fundamental representations with charge smaller than the observed $Q_F$ exist as
\begin{align}
\textrm{for}~&k\geq 0    &  2k&~\textrm{fund. reps}~\exists ~\textrm{with}~ |Q_i|<|Q_F| \\
\textrm{for}~&k<0    &  2|k|-1&~\textrm{fund. reps}~\exists ~\textrm{with}~ |Q_i|<|Q_F|
\end{align}
A few notes on the compositeness degree: $(i)$ $k$ can be negative since the spectrum of charges extends to negative values -an illustration of this case is given in the appendix- fig.~\ref{fig:negk}, $(ii)$ for the case of $U(1)$, negative $k$ values however can be mapped to positive $k$ (the lattice is symmetric under $Q\to-Q$), and we consider only $k\geq 0$ without loss of generality and $(iii)$ this is a group theoretical index, if $k\neq 0$, $F$ can be built out of smaller representations but it does not mean $F$ is necessarily a bound state. One has that possible charges for a fundamental representation for each $k$ are
\begin{align}
&&&U(1)\times SU(3) &&U(3)\\
    \frac{Q}{Q_F}&= & &\frac{n}{1+k}, & &\frac{1+3m}{1+3k},
\end{align}
with $n,m$ integers. The generating element for $Z_N$ depends on $k$, as can be obtained by rescaling the charge $Q_F$ in eq.~(\ref{xiN})
\begin{align}
    \xi_{(k)}=e^{2\pi i n_3/3}e^{-2\pi i (1+3k)Q/(3Q_F)}.
\end{align}
While the spectrum of charges gets ever more crowded for larger $m,l$, at finite order there are charges in $U(1)\times U(3)$ that cannot be `faked' by $U(3)$ and the minimum possible charge for each case and degree of compositeness is shown in tab.~\ref{tab:Qmin3}. On the other hand the limit $k\to \infty$ can be taken as a circle of ever larger radius that approaches $\mathbb{R}$ itself.

The culprit for the opening of this door to a vertiginous descent is of course the $U(1)$ factor; for $SU(N)$ the smallest representation is unambiguously identified, for $U(1)$ one cannot be sure just looking at the electric spectrum. One would need other type of observations to put a halt to this free fall; a possibility is evidence for a larger broken non-abelian group that contains $U(1)$ as done in grand unified theories, another would be the detection of a monopole so that Dirac's quantisation fixes the smallest possible charge. 

\begin{table}[h!]
    \centering
    \begin{tabular}{c|c|c|c|c|c|c|c}
    $G~\backslash$ cd($|k|$)& & 0&1 &2 & 3&$\dots$&$k$\\\hline
      $U(1)\times SU(3)$  & $Q_F/Q_{\rm min}$& 1 &2 &3 &4&&$1+k$\\
      $U(3)$  & $Q_F/Q_{\rm min}$& 1& 4&7 &10&\dots &$1+3k$\\
        & &&$-2$&$-5$&$-8$&&$1+3k$
    \end{tabular}
    \caption{Inverse possible minimum $U(1)$ charge for $U(1)\times SU(3)$ (first row) and $U(3)$ (last two rows) as a function of the compositeness degree (cd).}
    \label{tab:Qmin3}
\end{table}

A word on notation; the largest group with a given Lie algebra is the universal cover, the set of representations common to all possible groups is the electric root lattice and the spectrum of possible representations (Wilson lines) is correlated with the spectrum of 't Hooft lines~\cite{tHooft:1977nqb}. Indeed in the framework of generalised symmetries that has brought renewed attention to the global structure of groups and new ideas to phenomenology~\cite{Cordova:2022fhg,Cordova:2024ypu}, the centre is associated with an electric one-form discrete symmetry whereas taking the quotient makes a one-form magnetic symmetry emerge~\cite{Aharony:2013hda}. Equivalently, one can understand the quantisation of charge as the consequence of a discrete one form symmetry. Let us note that the explicit form of this symmetry depends on what we have termed compositeness degree.

{\bf The Standard Model group.} The case of the Standard Model starts from the universal cover, which is what is usually referred to by the SM group $\tilde G$ in eq.~(\ref{eqtG}),
which has centre $Z_6$ compatible with all SM representations. Taking the quotient by each of the possible subgroups yields $G_p$ 
\begin{align}
    G_p&\equiv SU(3)_c\times SU(2)_L \times U(1)_Y/Z_p\,,\\ \label{ZpElms}
    Z_p&=\left\{ \begin{array}{cc}
    Z_1& \xi_{(6)}\\
Z_2& \left\{1\,,\,\xi_{(3)}\right\} \\
Z_3& \left\{1\,,\,\xi_{(2)}\,,\,\xi_{(2)}^2\right\} \\
Z_6& \left\{1\,,\,\xi\,,\,\xi^2\,,\,\xi^3\,,\,\xi^4\,,\,\xi^5\right\}
    \end{array}
    \right.\,.
\end{align}
The fact that each of these cases is a distinct group is emphasized by identifying that
\begin{align}
G_6&=\tilde G/Z_6=S(U(3)\times U(2))\,,\\
G_3&=\tilde G/Z_3=U(3)\times SU(2)\,,\\
G_2&=\tilde G/Z_2=SU(3)\times U(2)\,.\\
G_1&=\tilde G/Z_1
\end{align}
Where by the quotient by $Z_1$ we mean the condition that $\xi_{(6)}$ be a representation of the identity.
These four possibilities were laid out in \cite{Hucks:1990nw} and more recently in the context of generalised symmetry in \citep{Tong:2017oea}.

The elements of the discrete groups do depend on the compositeness degree $k$, one has that for $k=0$ all groups can be given in terms of $\xi$~\cite{Tong:2017oea}, which is not the case in general
\begin{align}
    \xi&=e^{2\pi(1+6k) Q_Y i}\,e^{2\pi i n_c/3}\, e^{i\pi n_L} \label{xiZ6}\\
    \xi_{(2)}&= e^{4\pi (1+3k)Q_Y i}\,e^{4\pi i n_c/3}\label{xiZ3}\\
    \xi_{(3)}&=  e^{6\pi (1+2k)Q_Y i}\,e^{i\pi n_L}\label{xiZ2}\\
    \xi_{(6)}&= e^{12\pi (1+k) Q_Y i}\label{xiZ1}
\end{align}
where $n_c$, ($n_L$) is the n-ality under $SU(3)_c$ ($SU(2)_L$) of the representation $\xi_{(6/p)}$ acts on. The fact that there is a centre at all is nontrivial; take $k=0$, the procedure that lead to eq.~(\ref{xiN}) can be used to fix the $U(1)_Y$ factor in $\xi$ to make say the left handed quark doublet invariant under its action. It is then not guaranteed but is in fact the case that all other five SM representations are also invariant as can be corroborated with tab.~\ref{tab:chargesSM}.
 
The distinction between groups, for the same matter content, is a global one and they all yield the same perturbative dynamics. Paradoxical as it might seem, let us review now this perturbative dynamics for later use. The covariant derivative reads
\begin{align}
    D_\mu= &
    \partial_\mu+i g_i A^{i,a}_{\mu}\, T_{i,a}\,,
\end{align}
where $A$ stands for gauge bosons in the SM and $i$ runs through colour, isospin and hypercharge; $A_i=\left\{G,W,B\right\}$, $g_i=\left\{g_c,g,g_Y\right\}$, $T_i=\left\{T_{c},T_L,Q_Y\right\}$ where $Q_Y$ is the hypercharge operator with eigenvalues given in tab.~\ref{tab:chargesSM}. Our convention for the normalisation of generators in the fundamental representation is the usual
\begin{align}
    \textrm{tr}\left( T_{i,a}({\rm F})  T_{i,b}({\rm F})\right)&=\frac{\delta_{ab}}2\,,
\end{align}
that is $T_{c,a}=\lambda_a/2$, $T_{L,I}=\sigma_I/2$ with $\lambda$ ($\sigma$) the Gell-Mann (Pauli) matrices.
The field strength can be obtained from the commutator of derivatives as
\begin{align}
    [D_\mu,D_\nu]=ig_i F_{\mu\nu}^{i,a} \, T_{i,a}\,,
\end{align}

and the gauge boson EoM reads
\begin{align}D^\nu F_{\nu\mu}^{i,a}&=g_i J_\mu^{i,a}\,,  
\end{align}
where $g_i J_\mu^i=-\partial\mathcal  L/\partial  A^i_\mu$, and in particular
\begin{align}
     J^\mu_{i,a}=\sum_\psi\bar\psi \gamma^\mu  T_{i,a}\psi +i H^\dagger ( T_{i,a} D^\mu- \stackrel{\leftarrow}{D}^\mu  T_{i,a} )H \,.
\end{align}
It is also useful to note that the bosonic part of the current is, after EWSB,
\begin{align}\label{SBJY}
     J_Y^H&=\frac{v^2 Z(g/c_w)}{4}+(\textrm{field})^{p\geq2}\\\label{SBJL}
     J_{L,a}^H&=-\frac{v^2}{4}\left\{ \begin{array}{c}
         gW_{1,2}  \\
         g/c_w Z 
    \end{array}
    \right.+(\textrm{field})^{p\geq2}
\end{align}
where as usual $\tan(\theta_w)=g_Y/g$ and $Z=c_wW_3-s_wB$. 

\section{Spectrum for each group}\label{secSpct}
This section determines the possible spectrum of representations for each group. For illustration purposes this is first done for the simplest case, $k=0$, to later give the results for general $k$.
\subsection{Null compositeness degree}
 Let us first discuss the case of compositeness degree $k=0$.
The condition for taking the quotient by $Z_p$ translates into selecting matter representations insensitive (invariant) under the action of $Z_p$. Considering all elements in a specific $Z_p$ can be given by integer powers of a generating element (the first non-trivial term in eq.~\eqref{ZpElms}), it suffices that representations are invariant under the generating element. Explicitly the conditions on the spectrum for each group and a certain representation $R$, read
\begin{align}\label{Z6cond}
   G_6:&& \xi R&=R,
    & \frac{n_c}{3}+\frac{n_L}{2}+Q_Y=&\mathbb Z\,,\\
   G_3:&& \xi_{(2)} R&=R,
   & \frac{2n_c}{3}+2Q_Y=&\mathbb Z\,,\label{Z3cond}\\
   G_2:&& \xi_{(3)} R&=R,
   & \frac{n_L}{2}+3Q_Y=&\mathbb Z\,,\label{Z2cond}\\
   G_1:&& \xi_{(6)} R&=R, & 6Q_Y=&\mathbb Z\,.\label{Z1cond}
\end{align}

The case of $G_1$ where the zentrum is present and the quotient is the identity, as elaborated in sec.~\ref{SMGr} does still imply a constraint on the spectrum; indeed for $Z_6$ to be a group $\xi^6=\xi_{(6)}=1$ which is only true if charge is quantised in units of $Q_Y(q_L)$.

\begin{figure}
    \centering
\begin{tikzpicture}
\begin{scope}[xshift=-33cm, yshift=1.3cm]
\draw[gray,thick,dashed](3,1/2,0)--(0,1/2,0)--(0,1/2,3)--(3,1/2,3)--(3,1/2,0)--(3,0,0)--(3,0,3)--(0,0,3)--(0,1/2,3);
  \draw[gray,thick,dashed](3,1/2,3)--(3,0,3);
  \draw[gray,thick,dashed](3,0,0)--(6,0,0)--(6,0,3)--(3,0,3);
  \draw[gray,thick,dashed](6,0,3)--(6,1/2,3)--(3,1/2,3);
    \draw[gray,thick,dashed](6,0,0)--(6,1/2,0)--(6,1/2,3);
  \draw[gray,thick,dashed](3,0,0)--(0,0,0)--(0,1/2,0);
  \draw[gray,thick,dashed](0,0,0)--(0,0,3);  \draw[gray,thick,dashed](6,1/2,0)--(3,1/2,0);
\filldraw [red] (0,1/2,3) circle (2pt);
  \filldraw [black] (0,0,3) circle (2pt);
  \filldraw [red] (3,0,3) circle (2pt);
  \filldraw [red] (3,1/2,3) circle (2pt);
  \filldraw [black] (3,0.5,0) circle (2pt);
  \filldraw [red] (3,0,0) circle (2pt);
  \filldraw [red] (6,0,3) circle (2pt);
  \filldraw [red] (6,0.5,0) circle (2pt);
  \filldraw [red] (6,0,0) circle (2pt);
  \filldraw [red] (6,1/2,3) circle (2pt);
\filldraw [red] (0,0.5,0) circle (2pt);
\filldraw [red] (0,0,0) circle (2pt);
\draw (-0.35,1/2,3) circle node{$\frac{1}{6}$};
\draw (3.2,0.73,0) circle node{$\frac{1}{6}$};
  \draw(-0.5,-0.5,2.5) node{0};
  \draw(2.5,-0.5,2.5) node{1};
    \draw(5.5,-0.5,2.5) node{2};
    \draw(-0.8,2,2.5) node{\large $G_1$};
\end{scope}

\begin{scope}[xshift=-33cm, yshift=-2cm]
\draw[gray,thick,dashed](3,1,0)--(0,1,0)--(0,1,3)--(3,1,3)--(3,1,0)--(3,0,0)--(3,0,3)--(0,0,3)--(0,1,3);
  \draw[gray,thick,dashed](3,1,3)--(3,0,3);
  \draw[gray,thick,dashed](3,0,0)--(6,0,0)--(6,0,3)--(3,0,3);
  \draw[gray,thick,dashed](6,0,3)--(6,1,3)--(3,1,3);
    \draw[gray,thick,dashed](6,0,0)--(6,1,0)--(6,1,3);
  \draw[gray,thick,dashed](3,0,0)--(0,0,0)--(0,1,0);
  \draw[gray,thick,dashed](0,0,0)--(0,0,3);  \draw[gray,thick,dashed](6,1,0)--(3,1,0);
\filldraw [blue] (0,1,3) circle (2pt);
  \filldraw [black] (0,0,3) circle (2pt);
  \filldraw [blue] (3,0,3) circle (2pt);
  \filldraw [blue] (3,1,3) circle (2pt);
  \filldraw [black] (3,0.5,0) circle (2pt);
  \filldraw [blue] (6,0,3) circle (2pt);
  \filldraw [blue] (6,0.5,0) circle (2pt);
  \filldraw [black] (6,1,3) circle (2pt);
\filldraw [blue] (0,0.5,0) circle (2pt);
\draw (-0.35,1,3) circle node{$\frac{1}{3}$};
\draw (3.25,0.35,0) circle node{$\frac{1}{6}$};
  \draw(-0.5,-0.5,2.5) node{0};
  \draw(2.5,-0.5,2.5) node{1};
    \draw(5.5,-0.5,2.5) node{2};
    \draw(-0.8,2,2.5) node{\large $G_2$};
\end{scope}

\begin{scope}[xshift=-32cm, yshift=-10.5cm]
    \draw[thick,->] (-1,0,6) -- (1,0,6) node[anchor=north west] {$n_c$};
\draw[thick,->] (-1,0,6) -- (-1,0,3) node[anchor=south east] {$n_L$};
\draw[thick,->] (-1,0,6)--(-1,2,6) node[anchor=south west] {$Q_Y$};
\end{scope}

\begin{scope}[xshift=-33cm, yshift=-6cm]
\draw[gray,thick,dashed](3,1.5,0)--(0,1.5,0)--(0,1.5,3)--(3,1.5,3)--(3,1.5,0)--(3,0,0)--(3,0,3)--(0,0,3)--(0,1.5,3);
  \draw[gray,thick,dashed](3,1.5,3)--(3,0,3);
  \draw[gray,thick,dashed](3,0,0)--(6,0,0)--(6,0,3)--(3,0,3);
  \draw[gray,thick,dashed](6,0,3)--(6,1.5,3)--(3,1.5,3);
    \draw[gray,thick,dashed](6,0,0)--(6,1.5,0)--(6,1.5,3);
  \draw[gray,thick,dashed](3,0,0)--(0,0,0)--(0,1.5,0);
  \draw[gray,thick,dashed](0,0,0)--(0,0,3);  \draw[gray,thick,dashed](6,1.5,0)--(3,1.5,0);
  \filldraw [ForestGreen] (0,0,0) circle (2pt);
  \filldraw [black] (0,0,3) circle (2pt);
  \filldraw [black] (0,1.5,0) circle (2pt);
  \filldraw [ForestGreen] (3,0.5,3) circle (2pt);
  \filldraw [black] (4.15,1.7,3) circle (2pt);
  \filldraw [ForestGreen] (6,1,0) circle (2pt);
  \filldraw [black] (6,1,3) circle (2pt);
\filldraw [ForestGreen] (0,1.5,3) circle (2pt);
\draw (-0.75,1.25,2) circle node{$\frac{1}{2}$};
\draw (5.75,0.85,0) circle node{$\frac{1}{3}$};
\draw (2.65,0.5,3) circle node{$\frac{1}{6}$};
  \draw(-0.5,-0.5,2.5) node{0};
  \draw(2.5,-0.5,2.5) node{1};
    \draw(5.5,-0.5,2.5) node{2};
  \draw(-0.8,2.5,2.5) node{\large $G_3$};
\end{scope}

\begin{scope}[xshift=-33cm, yshift=-11cm]
\draw[gray,thick,dashed](3,3,0)--(0,3,0)--(0,3,3)--(3,3,3)--(3,3,0)--(3,0,0)--(3,0,3)--(0,0,3)--(0,3,3);
  \draw[gray,thick,dashed](3,3,3)--(3,0,3);
  \draw[gray,thick,dashed](3,0,0)--(6,0,0)--(6,0,3)--(3,0,3);
  \draw[gray,thick,dashed](6,0,3)--(6,3,3)--(3,3,3);
    \draw[gray,thick,dashed](6,0,0)--(6,3,0)--(6,3,3);
  \draw[gray,thick,dashed](3,0,0)--(0,0,0)--(0,3,0);
  \draw[gray,thick,dashed](0,0,0)--(0,0,3);  \draw[gray,thick,dashed](6,3,0)--(3,3,0);
\filldraw [black] (0,3,3) circle (2pt);
\filldraw [black] (0,1.5,0) circle (2pt);
\filldraw [black] (6,2.5,0) circle (2pt);
\filldraw [black] (3,0.5,0) circle (2pt);
\filldraw [black] (6,1,3) circle (2pt);
\filldraw [black] (3,2,3) circle (2pt);
\filldraw [black] (0,0,3) circle (2pt);
\draw (-0.75,2.5,2) circle node{$1$};
\draw (-0.25,1.25,0) circle node{$\frac{1}{2}$};
\draw(0.1,-0.25,0) node{1};
  \draw(-0.5,-0.5,2.5) node{0};
  \draw(2.5,-0.5,2.5) node{1};
    \draw(5.5,-0.5,2.5) node{2};
    \draw(5.75,2.25,0) node{$\frac{5}{6}$};
    \draw(3.35,0.5,0) node{$\frac{1}{6}$};
    \draw(5.5,0.5,2.5) node{$\frac{1}{3}$};
    \draw(1.5,0.75,0) node{$\frac{2}{3}$};
    \draw(-0.8,3.5,2.5) node{\large $G_6$};
\end{scope}
\end{tikzpicture}
    \caption{Constituent block of the lattice of allowed hypercharges and n-alities for each SM group $G_p$ and compositeness degree 0. The lattices are periodic in all three directions, with the period for hypercharge being $p/6$ for $G_p$. Representations allowed in $G_6$ are marked black and are common to all $G_p$, the rest are given in different colours. Note that given the periodicity of the $G_1$ lattice, every possible representation in $G_{p>1}$ is also allowed in $G_1$.}
    \label{fig:lattice-lines}
\end{figure}
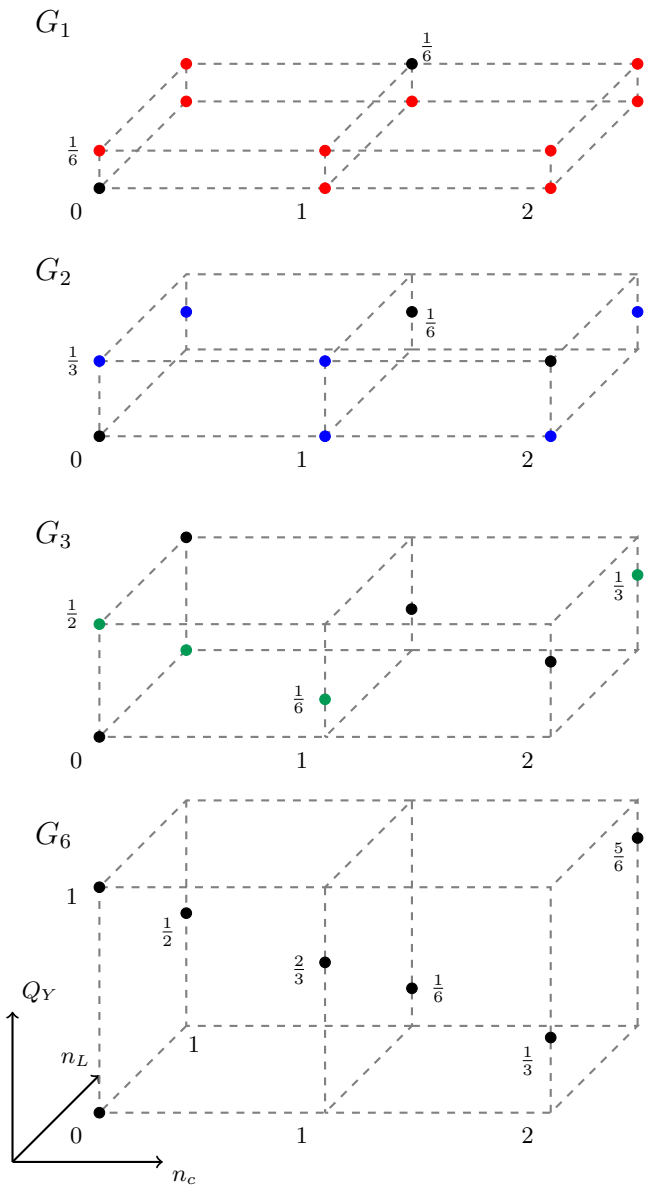

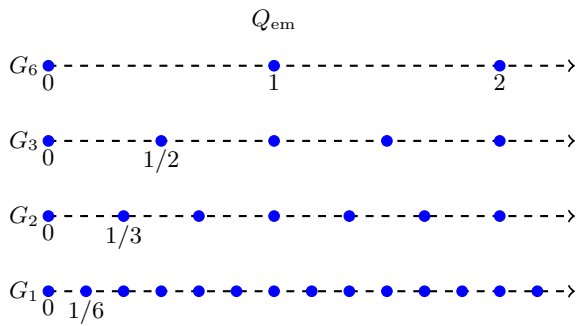
\begin{figure}
\begin{center}
    \begin{tikzpicture}
    \draw (3,3.6)node {$Q_{\rm em}$};
        \draw [thick,dashed,->] (0,3) node [anchor=east] {$G_6$} -- (7,3);
        \filldraw[blue] (0,3) circle (2pt) node [anchor=north] {\color{black}0};
        \filldraw[blue] (3,3) circle (2pt) node [anchor=north] {\color{black}1};
        \filldraw[blue] (6,3) circle (2pt) node [anchor=north] {\color{black}2};
        \draw [thick,dashed,->] (0,2) node [anchor=east] {$G_3$}-- (7,2);
        \filldraw[blue] (0,2) circle (2pt) node [anchor=north] {\color{black}0};
        \filldraw[blue] (1.5,2) circle (2pt) node [anchor=north] {\color{black}1/2};
        \filldraw[blue] (3,2) circle (2pt);
        \filldraw[blue] (4.5,2) circle (2pt);
        \filldraw[blue] (6,2) circle (2pt);
        \draw [thick,dashed,->] (0,1) node [anchor=east] {$G_2$} -- (7,1);
        \filldraw[blue] (0,1) circle (2pt) node [anchor=north] {\color{black}0};
        \filldraw[blue] (1,1) circle (2pt) node [anchor=north] {\color{black}1/3};
        \filldraw[blue] (2,1) circle (2pt);
        \filldraw[blue] (3,1) circle (2pt);
        \filldraw[blue] (4,1) circle (2pt);
        \filldraw[blue] (5,1) circle (2pt);
        \filldraw[blue] (6,1) circle (2pt);
        \draw [thick,dashed,->] (0,0) node [anchor=east] {$G_1$}-- (7,0);
        \filldraw[blue] (0,0) circle (2pt) node [anchor=north] {\color{black}0};
        \filldraw[blue] (1/2,0) circle (2pt) node [anchor=north] {\color{black}1/6};
        \filldraw[blue] (1,0) circle (2pt);
        \filldraw[blue] (3/2,0) circle (2pt);
        \filldraw[blue] (2,0) circle (2pt);
        \filldraw[blue] (5/2,0) circle (2pt);
        \filldraw[blue] (3,0) circle (2pt);
        \filldraw[blue] (7/2,0) circle (2pt);
        \filldraw[blue] (4,0) circle (2pt);
        \filldraw[blue] (9/2,0) circle (2pt);
        \filldraw[blue] (5,0) circle (2pt);
        \filldraw[blue] (11/2,0) circle (2pt);
        \filldraw[blue] (6,0) circle (2pt);
        \filldraw[blue] (13/2,0) circle (2pt);
    \end{tikzpicture}
\end{center}
\caption{Electric charge spectrum for hadrons and leptons in the case $k=0$.}\label{ElecQ}
\end{figure}

Each condition above leads to a discrete set of hypercharges, which can be visualised in a lattice as in fig.~\ref{fig:lattice-lines}. 
The possible representations will have electric charges given by
\begin{align}
    Q_{\textrm{em}}
    =T_{L,3}+Q_{Y}\,,
\end{align}
that is, the eigenvalue of the $SU(2)_L$ Cartan sub-algebra plus hypercharge.
The values for these eigenvalues are most accessible with the analogy to ordinary spin where $n_L$ signals the fermionic or bosonic character, i.e. whether the representation has semi integer or integer eigenvalues for any given spin direction respectively. The familiar result is, 
\begin{align}
    \begin{array}{c}T_{L,3}\\\textrm{Eigen.} \end{array}=\left\{ \begin{array}{ccc}
         n_L=1:\quad&
         \frac{m}{2}
         &
         m\in 2\mathbb{Z}+1\\
         n_L=0:& m
         & m\in\mathbb{Z} \\
    \end{array}
    \right.\,.
\end{align}
It is illustrative to use this input on the defining condition of the generating element of $Z_p$ to leave invariant any state, not necessarily elementary. Substituting hypercharge for electric charge and weak isospin, in the case of $G_6$ in eq.~(\ref{Z6cond}) as an example,
\begin{align}
   \mathbb{Z}= &\frac{n_c}{3}+Q_{\rm em}+\left(\frac{n_L}{2}-T_{L,3}\right)\,,\\
   \mathbb{Z} =&\frac{n_c}{3}+Q_{\rm em}+\left\{ \begin{array}{ccc}
        n_L=1: & \frac{1-m}{2} & m\in 2\mathbb{Z}+1 \\
        n_L=0: & -m & m\in \mathbb{Z}
    \end{array} \right.\,,
    \end{align}
    so weak isospin drops out of the equation to leave
    \begin{align}
G_6:& &    Q_{\rm em}=\mathbb{Z}-\frac{n_c}{3}\,.
\end{align}
Similar manipulations for $G_{p\leq 6}$ in eqs.~(\ref{Z3cond}-\ref{Z1cond}) lead to
\begin{align}
    G_3:&&Q_{\rm em}&=\frac12\mathbb{Z}-\frac{n_c}{3}\,,\\
    G_2:&&Q_{\rm em}&=\frac13\mathbb{Z}\,,\\
    G_1:&&Q_{\rm em}&=\frac16\mathbb{Z}\,.
\end{align}
The result of charges being $p/6$ for colour neutral states, i.e. leptons and hadrons, follows in each $G_p$. It is worth pointing out that this quantisation condition for charge makes it direct to derive the results of \cite{Tong:2017oea} for the charges of (colour and isospin neutral) monopoles from Dirac's quantisation condition $Q_{em}g_{em}=\mathbb Z$. Evidently, one obtains the results in table~\ref{tab:monop} while the spectrum of electric charges is depicted in figure~\ref{ElecQ}.
\begin{table}[h!]
    \centering
\begin{tabular}{r|c|c|c|c}
    &  $G_6$& $G_3$&  $G_2$&$G_1$\\ \hline
    $\textrm{min}[Q_{\rm em}\,(n_c=0)]$ &  $1$&  $1/2$&  $1/3$&$1/6$\\
    $\textrm{min}[g_{\rm em}\,(n_c=0)]$&  $1$&   $2$ & $3$&$6$
    \end{tabular}
    \caption{Allowed colour-neutral magnetic charges for $k=0$ for each group $G_p$.}
    \label{tab:monop}
\end{table}

\subsection{Arbitrary compositeness degree}

For general $k$ one has the forms of the different groups given in eqs.~(\ref{xiZ6}-\ref{xiZ1}) so that the procedure above follows giving the allowed representations in each case:
\begin{align}\label{Z6condG}
   G_6:&& \xi R&=R,
    & \frac{n_c}{3}+\frac{n_L}{2}+(1+6k)Q_Y=&\mathbb Z\,,\\
   G_3:&& \xi_{(2)} R&=R, 
   &\frac{2n_c}{3}+2(1+3k)Q_Y=&\mathbb Z\,,\label{Z3condG}\\
   G_2:&& \xi_{(3)} R&=R, 
   & \frac{n_L}{2}+3(1+2k)Q_Y=&\mathbb Z\,,\label{Z2condG}\\
   G_1:&& \xi_{(6)} R&=R, & 6(1+k)Q_Y=&\mathbb Z\,.\label{Z1condG}
\end{align}
while the electric spectrum is again independent of weak isospin and reads,
\begin{align}
G_6:& &    Q_{\rm em}&=\frac{1}{(1+6k)}\left(\mathbb{Z}-\frac{n_c}{3}\right)\,,\\
    G_3:&&Q_{\rm em}&=\frac{1}{2(1+3k)}\left(\mathbb{Z}-\frac{2n_c}{3}\right)\,,\\
    G_2:&&Q_{\rm em}&=\frac{1}{3(1+2k)}\mathbb{Z}\,,\\
    G_1:&&Q_{\rm em}&=\frac{1}{6(1+k)}\mathbb{Z}\,.
\end{align}
It is notable that the minimum possible charge for a lepton or hadron, shown in tab.~\ref{tab:Qmin6} has a value unique to each group and $k$.
\begin{table}[h!]
    \centering
    \begin{tabular}{c||c|c|c|c|c|c|c}
    $G_p\backslash$ cd& &0& 1&2 &3 & $\dots$&$|k|$\\\hline \hline
      $G_6:$  & $e/Q_{\rm min}$& 1& 7& 13 &19&&$1+6k$\\
        & && $-5$&$-11$ & $-17$&&\\
      $G_3:$  & $e/Q_{\rm min}$& 2& 8& 14 &20&\dots &$2+6k$\\
        & &&$-4$&$-10$ &$-16$&&\\
      $G_2:$  & $e/Q_{\rm min}$& 3& 9& 15 &21&\dots &$3+6k$\\ 
        &  &&$-3$ &$-9$&$-15$&&\\
      $G_1:$  & $e/Q_{\rm min}$& 6& 12&18 &24&\dots &$6+6k$\\
    \end{tabular}
    \caption{Inverse minimum electric charge for colour neutral states for each group and compositeness degree $k$.}
    \label{tab:Qmin6}
\end{table}

\section{All elemental quarks and leptons}\label{Sec:Smaller}

The SM matter content is compatible with all $G_p$ choices while e.g. $G_3$ allows for half-charge leptons and hadrons. One cannot build such half-charge particles in the SM, i.e. the full possible spectrum for $G_3$ is not spanned by SM representations. One is missing more elemental fields, where by elemental we mean the following: {\it a set of elemental fields spans, with the operation of representation combination and conjugate transform, all the possible charges in the  ($n_c,n_L,Q_Y$) lattice}.
As such a basis for elemental fields is not unique, it is natural to choose the basis with the smallest length (or dimension for the non-abelian part of the representation) for its vectors. Such a basis for $G_6$, with $k=0$, would consist of SM fields
\begin{align}\nonumber
    G_6:& & & \quad \bar e_R \qquad\quad\bar \ell_L \qquad\quad d_R\\ 
    & & & (1,1)_{1}\,,\, (1,2)_{1/2}\,,\,(3,1)_{-1/3}.
\end{align}
It is theoretically agreeable that in this case Nature would have chosen to start the particle puzzle with the smallest components, with the rest of the fermions $q_L$, $u_R$,  required by anomaly cancellation and the Higgs doublet $H$ for massive particles. 
As for the other group choices, bases for $k=0$ could be
\begin{align}\nonumber
    G_3:& & & \quad\Xi \qquad\quad\Lambda \qquad\quad \Omega\\ 
    & & & (1,1)_{1/2}\,,\, (1,2)_{0}\,,\,(3,1)_{1/6}\,, \label{G3Reps}\\\nonumber
    G_2:& & & \quad\Sigma \qquad\quad \Delta \qquad\quad \Theta\\ 
    & & & (1,1)_{1/3}\,,\, (1,2)_{1/6}\,,\,(3,1)_{0}\,, \label{G2Reps}\\\nonumber
       G_1: & & & \quad \Phi \qquad\quad \Lambda \qquad\quad \Theta\\ 
    & & & (1,1)_{1/6}\,,\, (1,2)_{0}\,,\,(3,1)_{0}\,. \label{G1Reps}
\end{align}
It should be noted nonetheless that as part of the freedom in basis choice one can also span all representations with a single fractional charge field and two SM fields. The case with $k\neq0$ would have bases with smaller hypercharge readily obtainable from eqs.~(\ref{Z6condG}-\ref{Z1condG}).

The discovery of one of these particles, or any of the other particles characteristic of a given group,  would narrow down the choice  of possible SM groups.
 At present no sign for new particles, despite the expectations, is currently established. In this work we take that to be the case due to extra particles being heavier than the scales probed in past laboratory experiments. In this circumstance, inferring the presence of such particles could be direct, via their production at the energy frontier at LHC, or indirect via low energy effects. Both cases are largely determined by the gauge group properties that define the new states.

 For concreteness in the remainder of this section let us focus on case of $k=0$, the generalisation to other $k$ is straightforward.

Other than gauge interactions, the new fractional-charge particles can couple to SM fields via invariant operators, akin to the yukawa terms in the SM. When considering the composition of such operators, one can organise them by the number of fractional fields and SM fields. Due to their fractional i.e. more elementary nature, these new fields cannot couple linearly, this result can be derived based just on electromagnetic invariance and n-ality conservation. Take a fractional state, say a $Q_{em}=1/3$ lepton and couple it to $n_{e,u,d}$ electron, up quark and down quark fields. Invariance demands
\begin{align}
     0&=\frac{1}{3}+n_{e}+\frac{2n_u}{3}-\frac{n_d}{3} \,,& n_u+n_d&=0\, \textrm{mod}\, 3\,,\\
     0&=\frac{1}{3}+n_e+\mathbb{Z}+n_u\,, & &
\end{align}
where in the second line we subbed the triality condition on the charge conservation condition to obtain a relation with no solution for $n_i\in\mathbb{Z}$.
If follows that an invariant operator requires at least 2 fractional-charge fields which is a result that we can also translate into the necessary stability of (at least the lightest of) these fractional-charge states. Indeed the equation above can be taken as the charge conservation condition that forbids the decay.

As for the couplings to one SM particle, possible combinations of operators are shown in tab.~\ref{tab:Ops} for the different groups.
\begin{table}[]
    \centering
    \begin{tabular}{c|c|c|c}
         & $G_3$& $G_2$& $G_1$\\ \hline 
         $q_L$ & $\Omega\Lambda$& $\Delta\Theta$& $\Phi \Lambda \Theta$\\ 
     $u_R$ & $\Omega\Xi$& $\Sigma\Sigma\Theta$& $\Phi \Phi \Phi \Phi \Theta$\\
     $d_R$ & $\bar{\Xi}\Omega$ & $\bar{\Sigma}\Theta$& $\bar{\Phi} \bar{\Phi}\Theta$ \\
     $\ell_L$ & $\bar{\Xi}\bar{\Lambda}$& $\bar{\Sigma}\bar{\Delta}$& $\Lambda \bar{\Phi} \bar{\Phi} \bar{\Phi}$\\
     $e_R$ & $\bar{\Xi}\bar{\Xi}$& $\bar{\Sigma}\bar{\Sigma}\bar{\Sigma}$& $\bar{\Phi} \bar{\Phi} \bar{\Phi} \bar{\Phi} \bar{\Phi} \bar{\Phi}$ \\
     $H$ & $\Xi\Lambda$ & $\Sigma\Delta$ & $\Lambda \Phi \Phi \Phi$
    \end{tabular}
    \caption{Minimal combination of one SM field and several fractional-charge fields that make up an invariant operator.} 
    \label{tab:Ops}
\end{table}     
Renormalisable couplings would either be of the Yukawa type or in an extended scalar potential depending on the spin of the new particles. Given the range of choices and unknowns, here we assume these other couplings are subleading to the gauge couplings and deffer their study to a more complete phenomenological exploration.
In addition to the assumption of gauge couplings dominating the interaction of the new states, we also assume that, if scalar, new states do not acquire a vacuum expectation value and do not partake in electroweak symmetry breaking. The case in which they do affect EWSB is nonetheless of interest and would fall in the class of non-decoupling physics with HEFT$\backslash$SMEFT low energy limit which we leave for future study. Lastly we consider arbitrary gauge group representations but restrict to spin $0$ and $1/2$ for the new states, in the case of fermions we take Dirac fermions with both chiralities on the same representation but relaxing this assumption would yield the same order of magnitude results. Anomaly cancellation is a relevant question which we do not address here, let us simply note that non-perturbative anomalies have been shown to cancel for these groups~\citep{Garcia-Etxebarria:2018ajm,Davighi:2019rcd,Wan:2019gqr,Davighi:2020bvi}.

The phenomenology of these particles will be studied both in the regime where they can and cannot be produced at the LHC. We find it useful in order to adopt a single theory approach to both cases to integrate the new particles out in the path integral so that the low (high) energy effects can be studied with the real (imaginary) resulting effective action. Consider a complex scalar, where in the following we will use $\Nm$ to refer to new fractional charge fields,
\begin{align}
    Z_\Nm[A]=\int\mathcal  D\phi \mathcal D\phi^\dagger e^{i S+ \delta \phi^\dagger S^{(2)} \delta\phi+\mathcal O(\delta\phi^3)}\,,\\
    =C e^{-c_{\textrm{s}}\textrm{Tr}\log(- S^{(2)}_{\textrm{s}})}(1+\mathcal O(\hbar))\,,\end{align}
    where s labels the spin, $\mathcal O(\hbar)$ signals 2-loop corrections, and for a scalar of mass $M$
\begin{align}
    -S^{(2)}_{0}&= -(\ell-iD)^2+M^2\,,& \textrm{Tr}&=\int \frac{d^4\ell d^4x}{(2\pi)^4}\underline{\mbox{tr}}\,,
\end{align}
with $\underline{\mbox{tr}}$ a trace over each gauge sector and $c_0=1$. The case of a fermion involves instead a grassmanian integral and the log trace of the Dirac operator which can be brought into the form above as
\begin{align}
    -S^{(2)}_{1/2}=-\gamma^\mu(\ell_\mu-iD_\mu)\gamma^\nu(\ell_\nu-iD_\nu)+M^2\,,
\end{align}
where we have dropped a term $i[\gamma D,M]$ assuming the effective mass to be a constant. Let us define the generating functional contribution from fractional-charge particles
\begin{align}
    i\Wf=\log (Z_\Nm[A]/Z_\Nm[0])\,,\label{eqW}
\end{align}
so that, in effect
\begin{align}
    \log(Z_\Nm[A])&= -c_{\textrm{s}}\textrm{Tr}\left(\log(-S^{(2)}_{\textrm{s}}[A])\right) \,,\\
    & c_{\textrm{s}}=\left\{
    \begin{array}{cc}
         \textrm{complex scalar} & 1 \\
         \textrm{fermion} & -1/2 
    \end{array}\right.\,.
\end{align}
The phenomenology of fractional-charge particles to be studied in the remainder of this section  can be all derived from $W_\Nm$, both in the case in which they can be produced directly~(\ref{secPhDirect}) and when their mass is too high for production~(\ref{secPhEFT}). 
\subsection{Production of fractional-charge particles}\label{secPhDirect}
Given the gauge group representation of the new states, production rates are determined up to the unkown mass $M$. The prediction for the partonic cross section will be derived here via the optical theorem rather than the otherwise straightforward direct computation. This route is chosen so that the connection between high and low energy is made more direct and for ease of manipulation of the group theory algebra.
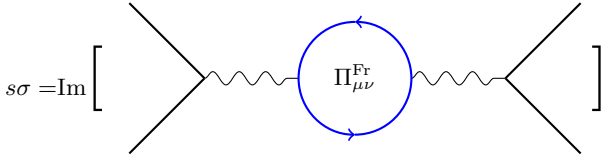
\begin{figure}
    \centering
\begin{center}
\begin{tikzpicture}
\draw (-3,0) node  {$s\sigma=$Im\Huge[};
    \draw[thick] (-2,1) --(-1,0);
    \draw[thick] (-2,-1) --(-1,0);
    \draw [decorate,decoration=snake] (-1,0) -- (0.25,0) 
    ;
    \draw [decorate,decoration=snake] (1.75,0)
    -- (3,0);
    \draw (1,0) node {$\Pi_{\mu\nu}^\Nm$};
    \draw[thick] (3,0) --(4,1);
    \draw[thick] (3,0) --(4,-1);
    \draw [thick,blue,->](1,0.75) arc (90:270:0.75);
    \draw [thick,blue,->](1,-0.75) arc (-90:90:0.75);
    \draw (4.25,0) node {\Huge]};
\end{tikzpicture}

\end{center}
\caption{The diagram in the effective action relevant for Drell-Yan production of new particles. The heavy mass limit of the real part of this diagram in turn determines certain operators in the EFT.}
    \label{fig:enter-label}
\end{figure}
In particular here we distinguish the cases of the new particles having colour, a case in which their production is dominated by $G+G\to \Nm+\bar\Nm$, and being colour neutral where instead production is approximated as Drell-Yan (DY). One can obtain the total cross section from the imaginary part of the effective action but also the differential cross section via Cutkovsky's rules as given in the appendix.
The partonic cross section in each case reads
\begin{align}\label{OpTh}
    \hat\sigma_{DY} &=  \left(\textrm{Im}\left( \Pf^{\mu\nu}\right)\frac{\mathcal M_\mu^{q\bar q\to A} \mathcal M_\nu^{A\to q\bar q}}{s^3}\right)_{\textrm{Fwd}}\,,
    \\
    \hat \sigma_{GG}&=\frac1s\textrm{Im}\left(\mathcal M_{GG\to GG}^\Nm\right)_{\textrm{Fwd}}\,,\end{align}
    where our convention for the invariant matrix element $\mathcal M$ in terms of the S-matrix is $S=1-i\mathcal M(2\pi)^4\delta^4(\Sigma p)$, Fwd stands for the forward limit, $\mathcal M_{GG}^{\Nm}$ is the contribution to the gluon four point amplitude with internal\Nm ~states, and $\Pf$ is
    \begin{align}
    &\frac{\delta^2\Wf[0]}{\delta  A_\mu(p)\delta A_\nu (k)}\equiv (2\pi)^4 \delta^4(p+k)\Pf^{\mu\nu}(p^2)\,.
\end{align}

Evaluation of each expression is carried out by taking the effective action of eq.~(\ref{eqW}) and the SM action in the path integral. This procedure does not require Feynman diagrams though it is straightforward to assign one to each term  in the computation. In the following let us outline the computation for DY with this method since we believe there is some value in doing so; the generating functional reads
\begin{align}\nonumber
    W_\Nm=&ic_{\rm s}\int\frac{d^4\ell d^4x}{(2\pi)^4}\underline{\mbox{tr}}\log\left(\frac{-(\ell-i\partial)^2+M^2-V(A)}{-(\ell-i\partial)^2+M^2}\right)\\
    =&ic_{\rm s}\int\frac{d^4\ell d^4x}{(2\pi)^2}\underline{\mbox{tr}}\sum_n\frac{-1}{n}\left(\frac{1}{M^2-(\ell-i\partial)^2}V(A)\right)^n,\label{logExpand}
\end{align}
where suppressing gauge group indices,
\begin{align}
    V(A)=\left\{(\ell-i\partial)_\mu,g A^\mu T\right\}+(g A_\mu  T)^2\,.
\end{align}
There is an ambiguity in the left or right operation of the free propagator, which can be solved expressing the logarithm as an integral of the inverse, see~\cite{Henning:2014wua}; here however conservation of momenta ensures the choice does not matter.
The contribution to eq.~(\ref{OpTh}) that will have an imaginary piece comes from the second term in the expansion of eq.~(\ref{logExpand}), and it reads (using that the variation wrt to the Fourier transform of $A$ is $\delta A(x)/\delta A(k)=e^{ikx}$) for a complex scalar
\begin{widetext}
\begin{align}\nonumber
    \frac{\delta^2W}{\delta A(k)_\mu^{i,a}\delta A(p)_\nu^{i,b}}&\supset
    \int \frac{d^4\ell d^4x}{2(2\pi)^4}
    \frac{-i}{M^2-(\ell-i\partial)^2}\underline{\mbox{tr}}\Big[\left\{(\ell-i\partial)_\mu , g_ie^{ipx} T_{i,a} \right\} 
   \times\frac{1}{M^2-(\ell-i\partial)^2}\left\{(\ell-i\partial)_{\nu}, g_ie^{ikx} T_{i,b} \right\}\Big]\\&+(\mu,a\leftrightarrow\nu,b)\\
   =&-ig_i^2\int d^4x \, e^{ix(p+k)}\int\frac{d^4\ell}{(2\pi)^4} \times\frac{(2\ell+p+2k)_\mu (2\ell+k)_\nu\underline{\mbox{tr}}\left(T_{i,a}T_{i,b}\right)}{(M^2-(\ell+p+k)^2)(M^2-(\ell+k)^2)}\,.
   \end{align}
\end{widetext}
   The integration over four space returns the momentum conservation Dirac delta so that
   \begin{align}
   \Pi^{\mu\nu}_{\Nm}=&-ig_i^2
   \int\frac{d^4\ell}{(2\pi)^4}\frac{(2\ell-p)_\mu(2\ell-p)_\nu\underline{\mbox{tr}}\left(T_a^i T_b^i\right)}{(\ell^2-M^2)((\ell-p)^2-M^2)}\,,\label{PiScalar}
\end{align}
whereas the fermionic result is
\begin{align}\nonumber
\Pi^{\mu\nu}_\Nm=&2ig_i^2\int\frac{d^4\ell}{(2\pi)^4} \\
   &\times\frac{\left[(2\ell-p)_\mu(2\ell-p)_\nu+p^2\eta_{\mu\nu}-p_\nu p_\mu\right]\underline{\mbox{tr}}\left( T_a^i T_b^i\right)}{(M^2-\ell^2)(M^2-(\ell-p)^2)}.\label{PiFermion}
\end{align}
The remaining trace is over all gauge indices, e.g. if the argument of the trace is a product of colour generators one has
\begin{align}
    \underline{\mbox{tr}}(T_{c,a} T_{c,b})=\Dim_L(R) \mbox{tr}(T_{c,a} T_{c,b})\,,
\end{align}
where $\Dim_L(R)$ is the dimension of the $SU(2)$ representation $R$. The remaining ordinary trace returns the Dynkin index, $\Dyn(R)$, which is a function of the representation $R$,
\begin{align}
    \textrm{tr}( T_{i,a}  T_{i,b})\equiv \Dyn_i({R})\delta_{ab}\,,
\end{align}
which is fully specified by our convention for the fundamental $\Dyn(F)=1/2$.

\begin{figure}
    \centering
    \includegraphics[width=.45\textwidth]{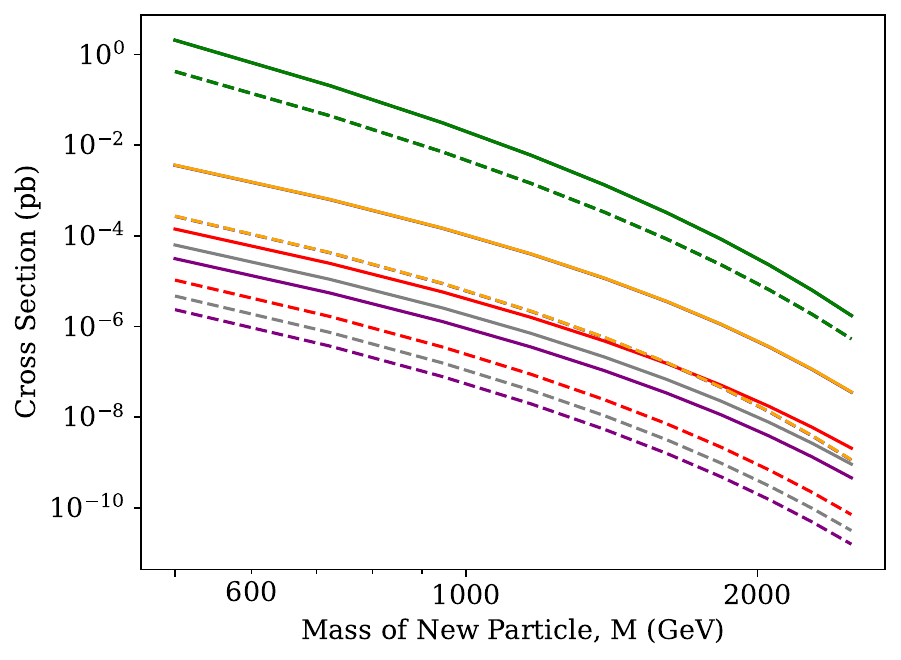}
    \caption{Hadronic cross section at LO against mass for the different representations in eqs.~(\ref{G3Reps}-\ref{G1Reps}) at the LHC at $13.5$~TeV. The pdfs are taken from \cite{Butterworth:2015oua}. The colours on this plot relate to the different representations as: $\Xi$  in red, $\Sigma$ in grey and $\Phi$ in purple. $\Delta,\Lambda$ are both in shown orange as the $\sigma_{qqL}$ contribution is significantly larger than $\sigma_{qqY}$. In green are both $\Theta,\Omega$ as $\sigma_{GG}$ again as the difference in hypercharge is insignificant. The line is solid if the representation is fermionic and dashed for scalar.}
    \label{fig:NPcrosssec}
\end{figure}

The imaginary part of the loop integral in eqs.~(\ref{PiScalar},\ref{PiFermion}) is a branch cut out of a logarithm, so that if one assembles the partonic cross section as obtained with the pdfs, the hadronic cross section for a proton-proton collision reads, for Drell-Yan mediated by $SU(2)_L$ bosons,
\begin{align}
     \sigma_{q\bar q,L}=&\frac{\pi\alpha_w^2\Dim_c\Dyn_L}{6
     s}\int \frac{dx dy}{xy}K_{\textrm s}(\hat s)\sum_a\textrm{Tr}\left[ T_{L}^a \mathbf{f}_q(x) T_{L}^a\mathbf{f}_{\bar q}(y)\right],
     \end{align}
     for hypercharge,
     \begin{align}
    \sigma_{q\bar q,Y}=&\frac{\pi\alpha_Y^2Q_{Y}^2\Dim_c\Dim_L}{6
    s}\int \frac{dx dy}{xy}K_{\textrm s}(\hat s)\sum_{i,\chi} Q_{Y,q_{\chi,i}}^2f_{q_i}f_{\bar q_i},
\end{align}
where $\hat s=xys$,  $\mathbf{f}_q=$Diag$(f_u,f_d)$ and $f_{q/\bar{q}}$ are the quark pdfs, $\chi$ stands for chirality, $\alpha_w=g^2/4\pi$, $\alpha_Y=g_Y^2/4\pi$ and the function $K$ is for  scalars and Dirac fermions,
    \begin{align}
         K_0(\hat s)&=\Theta(\beta^2)\frac{\beta^3(\hat s)}3\,, \\
         K_{1/2}(\hat s)&=\Theta(\beta^2)\left(2\beta(\hat s)-2\frac{\beta^3(\hat s)}{3}\right)\,,
    \end{align}
    with
    \begin{align}
    \beta(\hat s)=\sqrt{1-\frac{4M^2}{\hat s}}\,.
\end{align}
The case of a new fractional particle with colour reads instead
\begin{align}\nonumber
    \sigma_{GG}=&\frac{\pi\alpha_s^2\Dim_L4\Dyn_c^2}{s\Dim_c}\int    \frac{dxdy}{xy } f_G(x)f_G(y) L_{\rm s}(\hat s) \,,
    \end{align}
    with $f_G$ being the gluon pdf and the function $L_{\rm s}$ for a complex scalar and Dirac fermion read
    \begin{align}
    L_0=
    &\Theta(\beta^2) \Bigg[\beta\frac{2-\beta^2}{2}-\frac{1-\beta^4}2\textrm{atanh}(\beta)\\ \nonumber&\qquad +r\left(\beta\frac{3-5\beta^2}{24}-\frac{(1-\beta^2)^2}{4}\textrm{atanh}(\beta)\right) \Bigg],
\end{align}
\begin{align}
    L_{1/2}
    =&\Theta(\beta^2)\left[2\textrm{atanh}(\beta)-\frac{r\beta}{2}
    \right]-2L_0\,,
\end{align}
where
\begin{align}\label{rdef}
    \textrm{atanh}(x)&=\frac12\log\left(\frac{1+x}{1-x}\right), & 
    r&\equiv\frac{\Dyn_c(\textrm{Ad})\Dim_c(R)}{\Dyn_c(R)\Dim_c(\textrm{Ad})}\,.
\end{align}
The estimates for hadronic cross sections at the LHC are then, for the particles in eqs.~(\ref{G3Reps}-\ref{G1Reps}), shown in fig.~\ref{fig:NPcrosssec}.

Once produced, these particles would not decay into SM states only; they would either be stable or decay to other fractional and SM states. As such the characteristic signal of these particles is charged particle tracks in the detector, either leptonic or hadronic, quite unlike conventional BSM searches. The hadronic case seems more challenging to distinguish from background but still singular enough. Current analyses exist and e.g. rule out a particle $\Sigma(\Xi)$ with spin-1/2 and mass between 50~GeV and 60~GeV (600~GeV) at the 2$\sigma$ level~\cite{CMS:2024eyx}. Other results could be applied to particles in other representations, but here we do not aim at a comprehensive phenomenology, rather at a sketch of the main features that characterise $\Nm$ particles.
 On this note we close this section by noting that one such fractionally charged particle could  account for the excess in deposited energy $dE/dx$ observed at ATLAS~\cite{ATLAS:2022pib} with $3.3\sigma$ significance; this has been shown for $Q_{\rm em}>1$ in~\cite{Giudice:2022bpq} and we see no obstacle for the present $Q_{\rm em}<1$ case to also be accommodating of the excess.

\subsection{Low energy effects}\label{secPhEFT}

The focus of phenomenology is in fractional-charge representations which as discussed in sec.~\ref{Sec:Smaller} cannot couple to SM particles linearly. This implies, in the absence of SSB in the new sector, that the effects of these new particles at low energies start at one loop, as one can derive diagrammatically.
The effects at this order are given by equating the generating functional of eq.~(\ref{eqW}) to the effective action and expanding for large $M$; a universal formula is available, see e.g.~\cite{Henning:2014wua,Drozd:2015rsp}, and returns for a scalar field,
\begin{align}
    \nonumber\Gamma_{\rm EFT}^{\textrm{1-loop}}=&W_\Nm(D\ll M)= ic_{\textrm s}\textrm{Tr}(\log[ -S^{(2)}_s(D\ll M)])\\\nonumber=&\int \frac{c_sd^4x}{(4\pi)^2}\underline{\textrm{tr}}\Bigg[\frac{(D_\mu[D^\mu,D^\nu])^2}{60M^2}\\\nonumber& \qquad\qquad\quad-\frac{[D_\mu,D_\nu][D^\nu,D^\rho][D_\rho,D_\mu]}{90M^2}\Bigg]\\
    &+\mathcal{O}(M^{-4})\,.
\end{align}
The case of particles with spin follows the same steps and leads to
\begin{align}
    \mathcal L_{d=6}=-\frac{\underline{\textrm{tr}}\left(a_s(gD^\mu  F_{\mu\nu})^2-\frac{i\bar a_{3s}g^3 F_{\mu\nu}  F^{\nu\rho} F_{\rho\mu}}{3} \right)}{(4\pi)^2M^2120}\,,
\end{align}
we have for massive particles of spin $0$, $a_s=2$  and for spin $1/2$ Dirac particles, $a_s=8$. The last term will produce maximal helicity violating triple gauge couplings which are insensitive to the hypercharge of the new particle; as such in isolation they do not contain information about quantised charges so here concentrate on the first term above which reads, after using the EoM,
\begin{align}
    \mathcal L_{d=6}\supset-\frac{2\delta}{v^2} \left[a_c J_c^b J_c^b+a_L J_{L}^I J_{L}^I+a_Y J_{Y} J_{Y}\right]\,,
\end{align}
where
\begin{align}
    \delta&=\frac{a_s\Dim_L\Dim_cv^2}{(4\pi)^2240M^2}\,,& a_c&=\frac{\Dyn_cg_c^4}{\Dim_c}\,,\\
    a_L&=\frac{\Dyn_Lg^4}{\Dim_L} \,,& a_Y&=Q_Y^2(g_Y)^4\,.
\end{align}
The effect on $\mathcal L_{d=6,\text{QCD}}$ is to add four-fermion operators, but for $\mathcal L_{d=6,\text{EW}}$ after expanding the Higgs current as in eqs.~(\ref{SBJY}, \ref{SBJL}), one obtains also contributions to the masses and couplings of massive vector bosons; together with the SM terms they read,
\begin{align}
    \mathcal L\supset&-\frac{g}{\sqrt 2} W_+^\mu  J_{L,\mu}^-(1-\delta a_L)+h.c.\\&-\frac{g}{c_w} Z[(c_w^2-a_L\delta) J_{L,3}-(s_w^2-a_Y\delta) J_Y],
\end{align}
where $J_{L,-}=( J_{L,1}+i J_{L,2})$, and
\begin{align}
    M_W^2&=\frac{g^2v^2}{4}\left(1-a_L\delta\right)\,,\\
    M_Z^2&=\frac{g^2v^2}{4c_w^2}\left(1-a_L\delta-a_Y\delta\right)\,.
\end{align}
The ratio of $a_L$ and $a_Y$ takes a discrete set of values for $G_p$ that can be used to infer the quantum numbers of the new particle, however the overall scale of modifications is an unknown. To illustrate the possible correlation we first derive expressions for experimental inputs
\begin{align}
    G_F&=\frac{1}{\sqrt2v^2}(1-a_L\delta)\,, \\ 
    M_W&=\frac{gv}{2}(1-\frac{a_L\delta }{2})\,,\\
    \bar s_w&\equiv\frac{\sqrt{4\pi\alpha_{em}}}{2M_W(\sqrt{2}G_F)^{1/2}}=s_w(1+a_L\delta)\,,
\end{align}
which we substitute in the expression for two other observables that maximise the range of angles for the correlation of the Wilson coefficients
\begin{align}\label{rho3}
\rho_{\Gamma3}&\equiv\frac16\frac{M_Z^3\Gamma_W}{M_W^3\Gamma_Z^{\textrm{inv}}}=(1+\delta a_Y)\,,\\\label{rho5}
\rho_{\Gamma5}&\equiv\frac16\frac{(1-\bar s_w^2)M_Z^5\Gamma_W}{M_W^5\Gamma_Z^{\textrm{inv}}}=(1-2\bar t_w^2\delta a_L)\,.
\end{align}

\begin{figure}
    \centering
    \includegraphics[width=.35\textwidth]{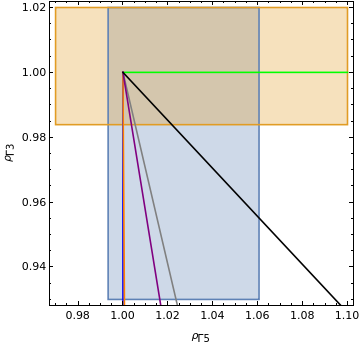}
    \caption{Correlation for deviations from the SM in the ratios of eqs.(\ref{rho3},\ref{rho5}). Green is for $SU(2)_L$ singlets (e.g. $\Phi,\Xi,\Sigma,\Omega,\Theta$, see eqs.~(\ref{G3Reps}-\ref{G1Reps})) blue is for hypercharge singlets in nontrivial $SU(2)_L$ reps (e.g. $\Lambda$ in eq.~(\ref{G3Reps}) orange is for $\Delta=(1,2)_{1/6}$, purple is for $(1,2)_{5/6}$, gray is for $(1,2)_{1}$ and black is for $(1,2)_2$. The bands show the experimentally allowed values, with the error dominated by $\Gamma_W$ and here given with 95\% CL taken from~\cite{TevatronElectroweakWorkingGroup:2010mao}.}
    \label{fig:CorrGa}
\end{figure}
One can make the correlation between observables $\delta$-independent as
\begin{align}
&\rho_{\Gamma3}=(1+r_{ew}(1-\rho_{\Gamma5}))\,,\\
&r_{ew}=\frac{\bar s_w^2\bar c_w^2 Q_Y^2 \Dim_L}{2\Dyn_L}\,.
\end{align}
In fig.~\ref{fig:CorrGa} we illustrate the correlation that would follow from each of the fractional-charge particles disscussed in sec.~\ref{Sec:Smaller}. The correlation shown here is not chosen for the observables with the best sensitivity, but rather as an illustrative example. In this sense it is worth noting that the correction to the ratio of electroweak gauge boson masses
\begin{align}
\rho_M&\equiv \frac{M_W^2}{(1-\bar s_w^2)M_Z^2}=1+\delta a_Y+2\bar t_w^2a_L\delta\,,
\end{align} 
  given $\delta\geq 0$, does have the right sign to account for an anomalously high $W$ mass compared with the $Z$ mass as reported by CDF~\cite{doi:10.1126/science.abk1781}; far less trivial would be to show this can be done consistently with other precision electroweak measurements. 

\section{Summary}\label{secSum}
The Lie algebra that characterises perturbatively gauge interactions in the Standard Model is shared by four groups $G_p=SU(3)_c\times SU(2)_L\times U(1)_Y/Z_p$ with $Z_p=\left\{Z_6,Z_3,Z_2,Z_1\right\}$. Each choice has a  characteristic possible lattice of representations with its most experimentally accessible property being the electric spectrum. For $G_p$ and degree $k$ of compositeness, $Q_Y(q_L)$ can be split into $|1+p\,k|$ representations and the minimum quantum for electric charge of hadrons and leptons is $p[6(1+p \,k)]^{-1}$, which reduces to $p/6$ if the hypercharge of $q_L$ is indivisible. If these fractionally charged particles exist, {\it(i)} the lightest of them would not decay and their production at the LHC would be signaled by fractionally charged heavy particle tracks, {\it(ii)} if lighter than the scale of inflation they would be present in a relic abundance and {\it(iii)} their effects at low energy, in the absence of SSB in the new sector, are loop suppressed and could be disentangled by ratios of Wilson coefficients. While these particles have received attention in the past, we believe their clear connection to the SM gauge group question calls for more systematic searches in the quest to answer one of the most fundamental questions in particle physics.
\\
\section*{Acknowledgements}
R. A., D. D., and M. W. are supported by the STFC under Grant No. ST/T001011/1. The authors are grateful to N. Lohitsiri and D. Tong for comments on the draft of this letter and stimulating discussions.
RA would like to thank S. Koren for bringing ref.~\cite{Hucks:1990nw} to our attention.

\bibliographystyle{unsrt}
\bibliography{sample}

\section{Appendix}\label{secApp}
\subsection{Negative compositeness degree}
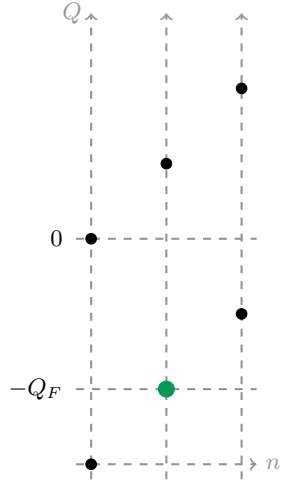
\begin{figure}[h!]
    \centering
     \begin{tikzpicture}
        \draw [Gray,thick, dashed,->] (-0.2,0) -- (2.2,0) node [anchor=west] {$n$};
        \draw [Gray,thick,dashed,->] (0,-0.2) -- (0,6) node [anchor=east] {$Q$};
        \draw [Gray,thick,dashed] (-0.2,1)  node [anchor=east] {\color{black}$-Q_F$\,} -- (2.2,1);
        \filldraw [black] (0,0) circle (2pt);
                \draw [Gray,thick,dashed] (-0.2,3) node [anchor=east] {\color{black} 0\,} -- (2.2,3); 
        \filldraw [black] (0,3) circle (2pt);
        \draw [Gray,thick,dashed,->] (1,-0.2) -- (1,6) ;
        \filldraw [ForestGreen] (1,1) circle (3pt);
       \filldraw [black] (1,4) circle (2pt);
        \draw [Gray,thick,dashed,->] (2,-0.2) -- (2,6) ;
        \filldraw [black] (2,2) circle (2pt);
        \filldraw [black] (2,5) circle (2pt);
    \end{tikzpicture}
    \caption{Negative compositeness degree example for $U(3)$ with the green entry being the observed representation and black entries other allowed representations.}
    \label{fig:negk}
\end{figure}

As an illustration of a negative compositeness degree consider the case in fig.~\ref{fig:negk} where the observed state $F$ happens to have charge a negative multiple of the elementary charge.
\subsection{Group theory invariants}

 One can classify the representations of $SU(2)$ by an integer $n$, the number of fundamental representations symmetrised to obtain the representation. One has
\begin{align}
    \Dim _L(n)&=n+1\,,& \Dyn_L(n)=&\frac{n(n+1)(n+2)}{3!\times 2}\,.
\end{align}
In terms of the familiar, spin, one can trade $n=2J$ for the familiar result $\Dim_2(n=2J)=2J +1$ and the Casimir $3\Dyn_2/\Dim_2=J(J+1)$. For $SU(3)$ one needs two integers for the number of fundamental $n$ and anti-fundamental $m$ representations that build up $R$.
Freudenthal's formula gives
\begin{align}
    \Dim_c(n,m)&=\frac{(m+1)(n+1)(n+m+2)}{2}\,,\\
    \frac{\Dyn_c(m,n)}{\Dim_c(n,m)}&=\frac{(m^3+n^3+3(m+n)+m n)}{4!}\,.
\end{align}
In terms of these one has $n_L=n$ mod $2$ and $n_c=n-m$ mod $3$. 
The literature often quotes results instead of for arbitrary representations, for arbitrary $N$ in $SU(N)$, the relation is $\Dim_N(\textrm{Ad})=N^2-1$, $\Dim_N(F)=N$, $\Dyn(F)=1/2$, $\Dyn($Ad$)=N$. Another often times used invariant is the Casimir $C$ related to the Dynkin index as:
\begin{align}
\sum_a T_a(R) T_a(R)\equiv&C(R)\mathbbm{1}\,, \\
 C(R)\Dim(R)=&\Dyn(R)\Dim(\textrm{Ad})\,.
\end{align}

\subsection{Differential cross sections}
The differential form of cross sections can be obtained with Cutkovsky's rules; the expressions of the functions $K,L$ are, for scalars
    \begin{align}
        K_0&= \Theta(\beta^2)\int\frac{dt_1}{s} 2\left(\frac{t_1u_1}{s^2}-\frac{M^2}{s}\right)\\
&\equiv\Theta(\beta^2)\int\frac{dt_1}{s}\bar\omega(t_1)        \,.
        \end{align}
        where  $t_1=t-M^2$, $u_1=u-M^2$ and $r$ is the group invariant given in eq.~(\ref{rdef}).
        For Dirac fermions
        \begin{align}
        K_{1/2}&= 2\Theta(\beta^2)\int\frac{dt_1}{s} \left(1-2\left(\frac{t_1u_1}{s^2}-\frac{M^2}{s}\right)\right)\\
        &=\int \frac{dt_1}{s}2(1-\bar\omega(t_1)).
    \end{align}
    For $G+G\to \Nm+\bar\Nm$ with scalars

\begin{align}
L_0&= \Theta(\beta^2)\int \frac{dt_1}{s}\left(1-\frac{u_1t_1}{s^2}r\right)\omega(t_1)\,,\\
\omega(t_1)&=\frac12-\frac{M^2s}{t_1u_1}\left(1-\frac{M^2s}{t_1u_1}\right)\,,
\end{align}
and for Dirac fermions

\begin{align}
L_{1/2}=2 \Theta(\beta^2)\int \frac{dt_1}{s}\left(1-\frac{u_1t_1}{s^2}r\right)\left(\frac{s^2}{4t_1u_1}-\omega(t_1)\right)\,. 
\end{align}
It is interesting to note that the positive definite nature of cross section demands $r\leq 4$ which is of course in agreement with $SU(N)$ where the largest $r$ is for the fundamental with $r(F)=2N^2/(N^2-1)$.

\end{document}